\documentclass{iopart}

\pdfoutput=1 
 \expandafter\let\csname equation*\endcsname\relax
  \expandafter\let\csname endequation*\endcsname\relax

\usepackage{amsmath}
\usepackage{amsthm,amssymb}
\usepackage[a4paper,left=2.0cm,right=2cm,top=2cm,bottom=2cm]{geometry}
\usepackage{graphicx} 
\graphicspath{{figs/}}

\usepackage{overpic}

\newcommand{\Ref}[1]{(\ref{#1})}

\newcommand{\nn}{\nonumber \\}

\numberwithin{equation}{section}

\begin{document}
\title{Analysis of mean cluster size in directed compact percolation near a damp wall}
\author{H Lonsdale$^1$, I Jensen$^1$, J W Essam$^2$, and  A L Owczarek$^1$}
\address{    $^1$ Department of Mathematics and Statistics,\\
    The University of Melbourne, Victoria~3010, Australia.\\
    \texttt{h.lonsdale@ms.unimelb.edu.au,
           i.jensen@ms.unimelb.edu.au, 	
     		a.owczarek@ms.unimelb.edu.au}
         \\[1ex]    
    $^2$ Department of Mathematics,\\
    Royal Holloway, University of London, Egham Surrey TW20 0EX, UK.\\
    \texttt{j.essam@rhul.ac.uk}
}


\begin{abstract}
We investigate the behaviour of the mean size of directed compact percolation clusters near a damp wall in the low-density region,
where sites in the bulk are wet (occupied) with probability $p$ while sites on the wall are wet with probability $p_w$. 
Methods used to find the exact solution for the dry case ($p_w=0$) and the wet case ($p_w=1$) turn out to be inadequate
for the damp case. Instead  we use a series expansion for the $p_w=2p$ case to obtain a second order inhomogeneous 
differential equation satisfied by the mean size, which exhibits a critical exponent $\gamma=2$, in common with the wet wall result. 
For the more general case of $p_w=rp$, with $r$ rational, we  use a modular arithmetic method of finding ODEs and
obtain a fourth order homogeneous ODE satisfied by the series. The ODE is expressed exactly in terms of $r$.
We find that in the damp region $0<r<2$ the critical exponent $\gamma^{\rm damp}=1$, in common with the dry wall result.

\end{abstract}

\section{Introduction} \label{sec intro}

Directed compact percolation,  introduced by Domany and Kinzel~\cite{DomanyKinzel}, is an exactly solvable model. 
The results for various cluster properties in the {\it bulk}\/ case, away from any confining walls, are given in~\cite{DomanyKinzel} and~\cite{Essam89}.
The addition of a {\it wet } wall to the model was considered in~\cite{EssamTan94}, and it was found that the critical exponents 
for the cluster properties follow that of the bulk case. However, the addition of a {\it dry} or non-conducting wall, 
considered in~\cite{BidauxPrivman, Lin, EssamGutt95, BrakEssam99}, produced different exponents to the wet and bulk cases.
This led to the consideration of a {\it damp} wall, introduced in~\cite{Lonsdale09}, which interpolates between the wet and dry cases.  
It was found that the critical behaviour followed the dry case, and the calculation of several cluster properties~\cite{EssamDomb, Lonsdale11} 
was possible using the same methods as near a dry wall.   

This paper extends the work on directed compact percolation near a damp wall, to consider the mean size of finite clusters in the low-density region. 
In previous work on the bulk~\cite{Essam89}, wet~\cite{EssamTan94} and low-density dry~\cite{EssamGutt95}  cases, the mean size was found 
by solving  the associated recurrence relations.  For other cluster properties near a damp wall --- percolation probability~\cite{Lonsdale09}, 
mean length~\cite{EssamDomb} and mean number of contacts~\cite{Lonsdale11} ---  the same methods  yielded a solution, 
albeit in a more complicated form, exhibiting the same critical behaviour as the dry case. 
So we proceed with the mean size near a damp wall, guided at first by the work near a dry wall in~\cite{EssamGutt95}. 
 
However, we find that the recurrence relations for the mean size cannot be solved using the same methods as for the dry wall case, 
and it can be shown that they do not have the same form of solution. 
This was supported by a functional equations approach, which led us to consider alternative methods of analysing the mean size. 

The series expansion  for the special case $p_w=2p$, which tends to a wet wall near the critical point,  can be successfully analysed using the Guess.m package \cite{guess} for Mathematica; 
 we find that the mean size in this case satisfies a second order inhomogeneous differential equation. 
We then applied a more involved series analysis method \cite{BGHJ_2008, Jensen2010}, which makes use of modular arithmetic to 
more efficiently find differential equations satisfied by the series for the mean size. In the general case $p_w=rp$ with $r$ rational
 the series is a solution to a homogeneous ODE of order 4 and degree 33, with the exception of the special cases corresponding to simpler models.
Analysis of the ODE shows that the critical exponent for the mean size in the general damp case ($0<r<2$) is $\gamma^{\rm damp}=1$
and the physical critical point occurs at $p_c=1/2$, in line with the dry result.

\subsection{Model}

The model of compact percolation is defined on a directed square lattice, the sites of which are the points in the ($t, x$)-plane 
with integer co-ordinates such that $t\ge 0, \ x\ge 1$,  and $t+x$\/ is even.  The damp wall is represented by the sites at $x=1$, 
where each wall site is `wet' (occupied) with probability $p_w$ and `dry' (unoccupied) with probability $q_w=1-p_w$, while
sites in the bulk (away from the wall) are wet  with probability $p$ and dry with probability $q=1-p$.
We begin with an initial {\it seed}\/ of $m$\/ contiguous sites at $t=0$, the midpoint of which is located $y$ units above the wall. 
The seed is placed with certainty, and a cluster is grown from this column by column according to the rules of directed compact percolation.
The new site $(t,x)$ becomes wet with certainty if both the previous sites $(t-1,x\pm 1)$ are wet. If only one of the previous sites is wet
the new site is wet with probability $p$ and dry with probability $q=1-p$. When both previous sites are dry the new site  is 
dry with certainty, thus ensuring that the cluster remains compact. 

\begin{figure}[htbp]
\begin{center}
 \includegraphics[scale=0.4]{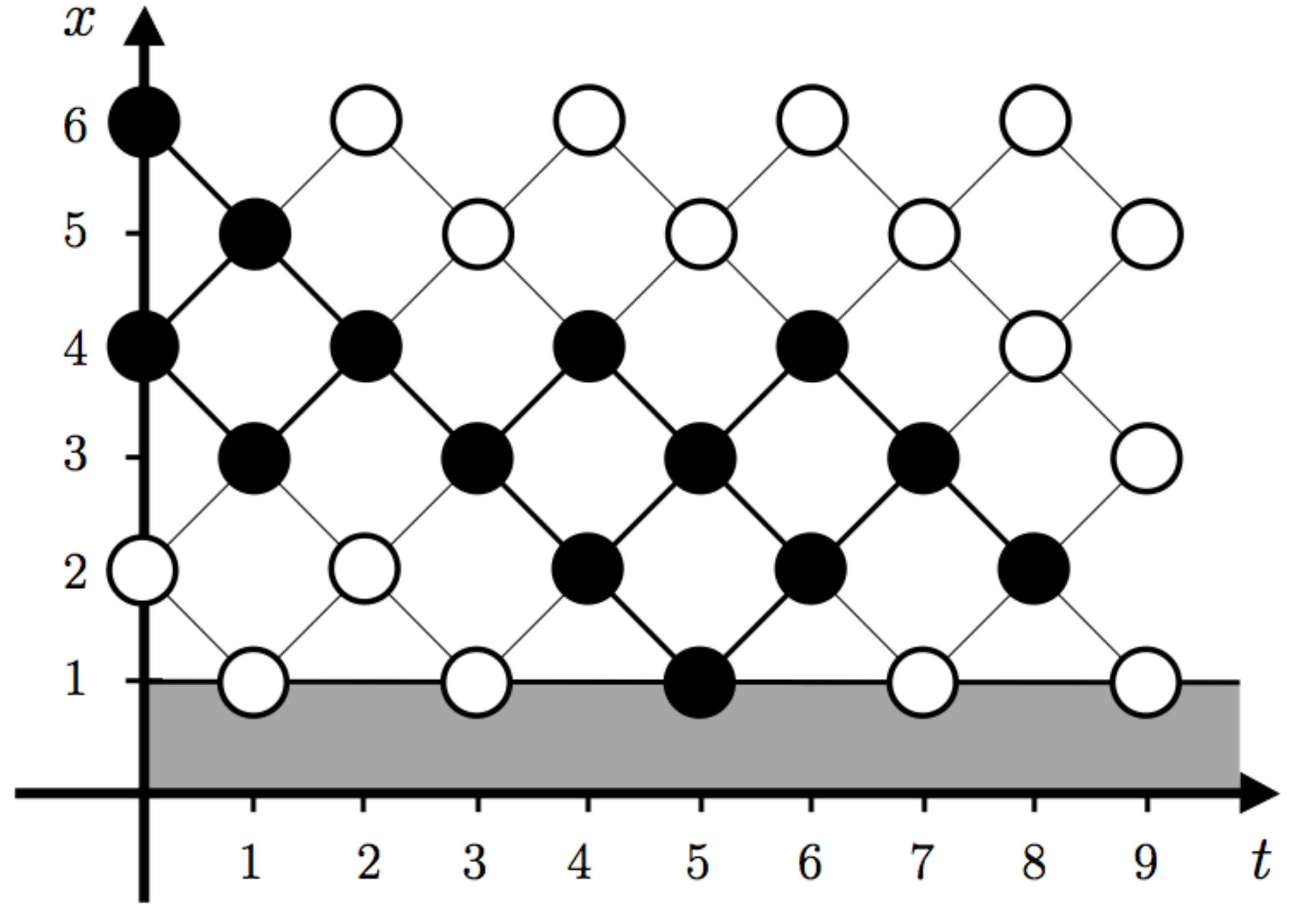}
\caption{ An example cluster, of size 14, grown from a seed of width $m=2$ with midpoint located $y=4$ units from the wall at $x=1$.
The probability of this cluster being grown from the seed can be calculated a column at a time, as 
 $(1)(pq)(q^2)(pq)(p^2)(p_wq)(p)(qq_w)(pq)(qq_w)=p^6q^8p_wq_w^2$. 
}
\label{eg_general}
\end{center}
\end{figure}

\subsection{Mean cluster size}
The size of a cluster is defined as the number of wet sites in the cluster, including the seed.  We will also consider adjacent wet sites on the wall to form part of the cluster.
We define $S_{m,y}(p,p_w)$ to be the mean size of finite clusters grown from a seed of width $m$ with midpoint $y$ units from the wall,
 and $\bar{S}_{m,y}$ to be the unnormalized mean size,
\begin{align}
	\bar{S}_{m,y}=S_{m,y}(p,p_w)Q_{m,y}\,,
\end{align} 
where $Q_{m,y}$ is the probability that a finite cluster is grown from this  seed. 
We note that in the {\it low-density} region, below the critical point $p=p_c$, we have $Q_{m,y}=1$ and hence
  $S_{m,y}=\bar{S}_{m,y}$. 
  
The critical exponent for the mean size,  $\gamma$, describes the behaviour near the critical point. As $p\to p_c$ we have
  \begin{align}
S_{m,y} \sim |p_c-p|^{-\gamma}\,.
 \end{align}
  We briefly review the results found for the mean size in other cases of directed compact percolation, which will guide our work on the damp wall case.

\subsubsection{Bulk case:}
The mean size of finite clusters in the bulk case, found in \cite{Essam89} by solving the associated recurrence relations, is 
\begin{align}
	S_m^{\rm bulk}(p)
	= 	\left\{ 
		\begin{matrix}
			\dfrac{m}{1-2p} \left(\dfrac{(1-p)^2}{1-2p}+\dfrac{m-1}2\right)  & \mbox{ for } p<\frac{1}2\,; \\
			\dfrac{m}{2p-1} \left(\dfrac{p^2}{2p-1}+\dfrac{m-1}2\right) & \mbox{ for } p>\frac{1}2\,;
		\end{matrix}
		\right.
\label{Smbulk}
\end{align}
and so the critical exponent $\gamma^{\rm bulk}=2$.

\subsubsection{Wet wall:}
The mean cluster size near a wet wall was  found in  \cite{EssamTan94}, again by solving the associated recurrence relations. 
This was done for clusters grown from a seed of width $m$ adjacent to the wet wall (the cluster then remaining attached due to the attractive wet wall) with the result 
\begin{align}
	2S_m^{\rm wet}(p)&=\frac{m-2p(1-p)}{(1-2p)^2}+\frac{2m^2-m}{|1-2p|}\,.
	\label{Swet}
\end{align}
This exhibits the same critical behaviour as the bulk case with $\gamma^{\rm wet}=2$. 

Note that the work in \cite{EssamTan94} does not include adjacent wall sites in a cluster's size, unlike the general damp model. This effectively shifts the location of the wall by one unit, and so we consider the result in \Ref{Swet} to apply to seeds beginning {\it on} the wall. 
We consider in particular the $m=1$ case,
\begin{align*}
  S_1^{\rm wet}(p) := S_{1,0}(p,1)  = \frac{(1-p)^2}{(1-2p)^2}.
\end{align*}

\subsubsection{Dry wall:}
In \cite{EssamGutt95}, the mean size of clusters near a dry wall was calculated in the low-density region by solving the recurrence relations. 
The result for the mean size of a cluster grown from a seed of width $m$, with midpoint $y$ units from a dry wall, is 
 \begin{align}
	S_{m,y}^{\rm dry}(p)
		& = S_m^{\rm bulk}(p)-\frac{mp^2}{(1-2p)^2}\left( \dfrac p{1-p}\right)^{y-m-1}, \ p<\frac{1}2 \label{Sdrybulk} \\
		& = \ \frac{m(m+1)}{2(1-2p)}+\frac{mp^2}{(1-2p)^2}\left(1-\left(\frac p{1-p}\right)^{y-m-1} \right), \ p<\frac{1}2\,.
\label{Sdrybulk2}
\end{align}
It can be seen from \Ref{Sdrybulk} that in the bulk limit, as $y\to \infty$, the dry case tends to the bulk result, 
and so in the bulk limit this expression has exponent $\gamma=2$. However, everywhere else 
the exponent is $\gamma^{\rm dry}=1$, as a factor of $(1-2p)$ cancels in the second term of \Ref{Sdrybulk2} for any integer value of $y$, and hence the dry wall mean size exhibits 
different critical behaviour to the bulk and wet cases.

We focus in particular on the case where a seed of a single site is situated adjacent to the wall, and we define $S(p)$ to be the mean size in this situation,
\begin{align}
	S^{\rm dry}(p) := \ S^{\rm dry}_{1,1}(p) \ = \ & \frac{1-p\,}{1-2p}\,, \quad p<\frac{1}2\,.
\label{Sdry11}
\end{align}

\section{Mean size near a damp wall} \label{meansize}

\subsection{Recurrence relations}
We set up the recurrence relations by considering the possibilities after one time step for a seed of width $m$, with midpoint $y$ units from the wall, for each of three seed classifications:  seeds located on, adjacent to, and away from the wall.

\subsubsection{Away from the wall,  \boldmath  $y>m$:}
In the bulk the cluster is unaffected by the wall. The corresponding recurrence for the mean size 
will therefore be the same as in the dry wall case \cite{EssamGutt95},
\begin{align}
	\bar{S}_{m,y}=pq\bar{S}_{m,y+1}+pq\bar{S}_{m,y-1}+p^2\bar{S}_{m+1,y}+q^2\bar{S}_{m-1,y}+mQ_{m,y}\,, \quad  y > m > 1\,.
\label{Sbulk}
\end{align}
This encompasses the four different possible configurations of the cluster in the column following the seed as illustrated in Figure \ref{gen_rec_size}. 
However, we note that for a seed consisting of a single site, that is $m=1$, the term with coefficient $q^2$ in \Ref{Sbulk} would not be present, 
as this would correspond to the cluster terminating. So we consider this case separately and impose the condition
\begin{align}
	\bar{S}_{1,y}=pq\bar{S}_{1,y+1}+pq\bar{S}_{1,y-1}+p^2\bar{S}_{2,y}+Q_{1,y}\,, \quad y > 1\,.
\label{Sbulk1}
\end{align}

\begin{table}[htdp]
\begin{center}
	\begin{tabular}{|l|c|c|c|c|}
		 \hline
		possible configurations: & a) \qquad \qquad  & b) \qquad \qquad & c) \qquad \qquad & d) \qquad \qquad \\
		& \includegraphics[scale=0.2]{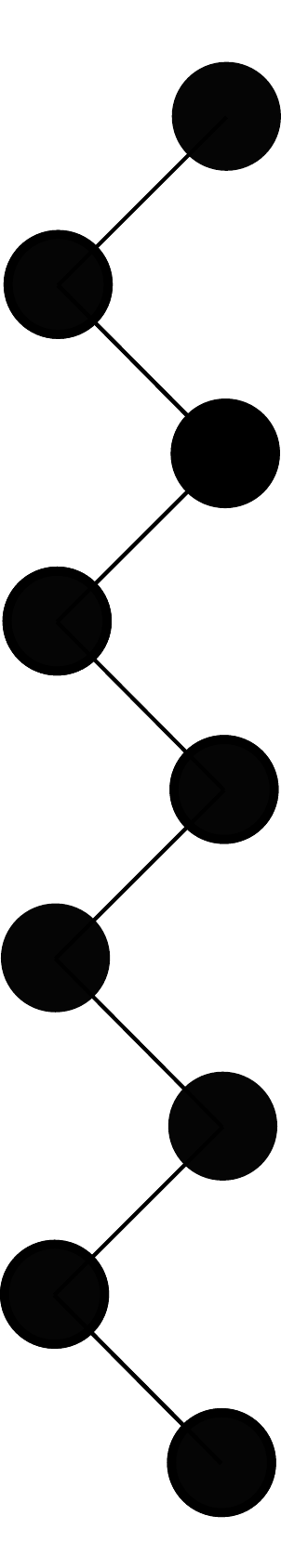}   &  \includegraphics[scale=0.2]{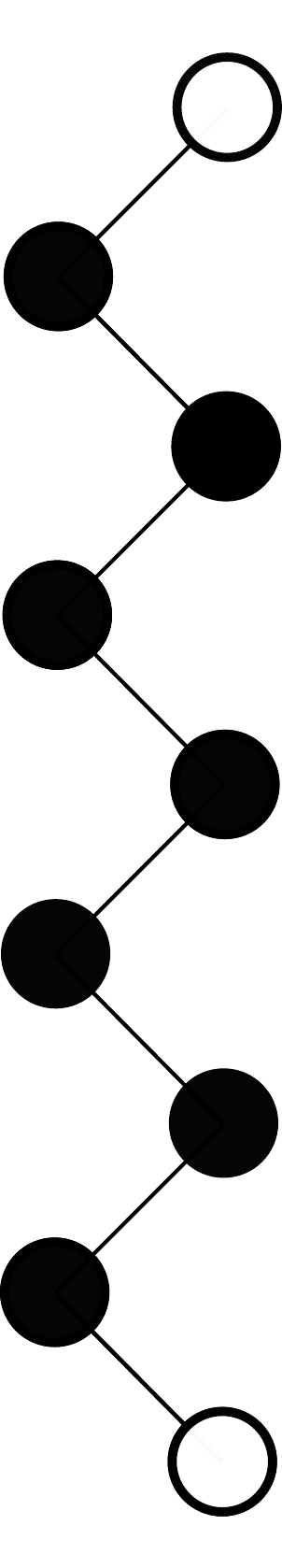}    
		&   \includegraphics[scale=0.2]{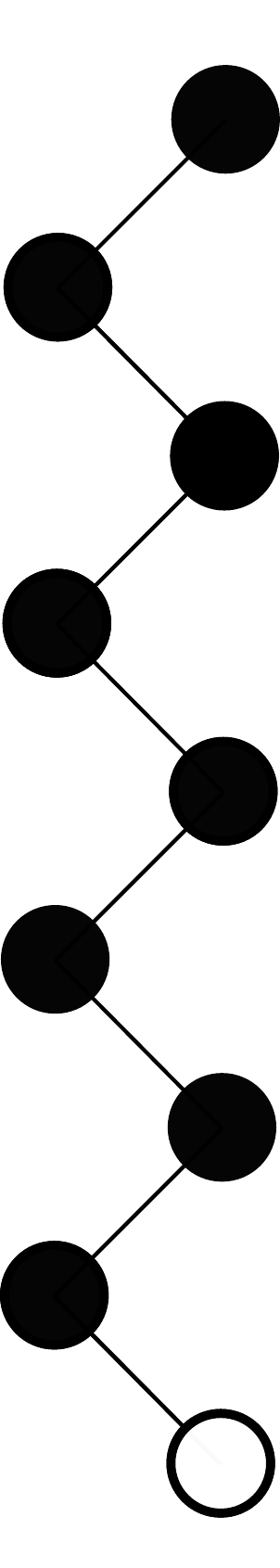}   &   \includegraphics[scale=0.2]{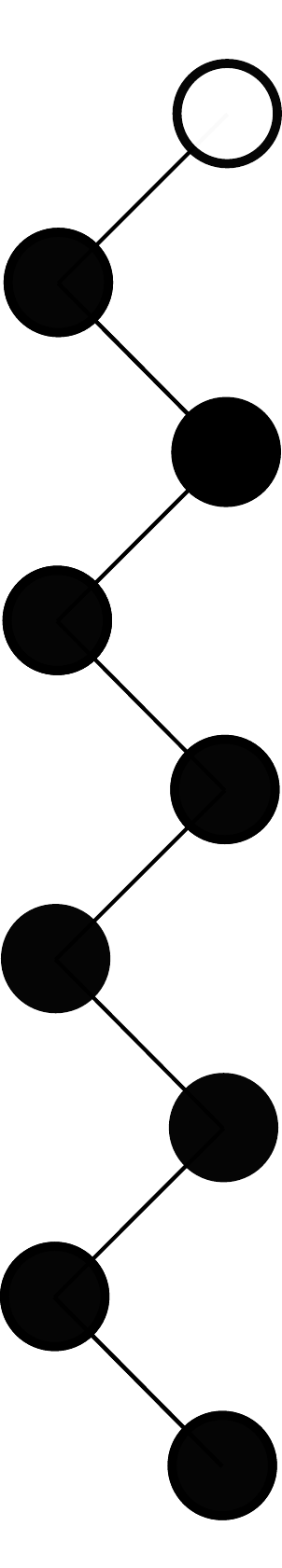}   \\
		 \hline
		\underline{\it $2^{nd}$ column, $t=1$} & & & &  \\
		cluster width: & $m+1$ & $m-1$ & $m$ & $m$ \\
		midpoint distance: & $y$ & $y$ & $y+1$ & $y-1$ \\
		\hline
		\underline{probability} & & & &  \\
		away from wall: &  $p^2$  & $q^2$ &  $p q$   &   $pq $  \\
		adjacent to wall: &  $pp_w$  & $qq_w$ &  $p q_w$   &   $qp_w $  \\
		on the wall: &  ---  & $q$ &  $p$   &   ---  \\
		 \hline
	\end{tabular}
	 \caption[Cluster configurations near a damp wall]{
	The different configurations possible for a cluster beginning with a seed of width $m$, with midpoint $y$ units from the wall, and their 
	probabilities for each seed classification, shown  through a sample cluster of initial seed width $m=4$.
	}\label{gen_rec_size}
\end{center}
\end{table}

\subsubsection{Adjacent to the wall,  \boldmath $y=m$:}
For a cluster having a seed adjacent to the wall, which corresponds to $y=m$, we can simply alter \Ref{Sbulk} to account for the probability $p_w$\/ that the adjacent wall site is wet, or dry with probability $q_w$\,, in place of an adjacent site in the bulk. 
Thus we have, for clusters adjacent to the damp wall,
\begin{align}
	\bar{S}_{m,m}=pq_w \bar{S}_{m,m+1}+p_wq\bar{S}_{m,m-1}+pp_w\bar{S}_{m+1,m}+qq_w\bar{S}_{m-1,m}+mQ_{m,m}\,, \quad m >1
\label{Swallpw}
\end{align}
Similar to \Ref{Sbulk1}, we consider separately a seed of width 1 adjacent to the wall, for which the mean size will satisfy
\begin{align}
	\bar{S}_{1,1}=pq_w\bar{S}_{1,2}+p_wq\bar{S}_{1,0}+pp_w\bar{S}_{2,1}+Q_{1,1} \,.
\label{Swallpw1}
\end{align}

\subsubsection{On the wall,  \boldmath $y=m-1$:}
If the seed includes a site on the wall,  which corresponds to $y=m-1$, then the cluster is unable to propagate downwards; 
so we can simply  focus on the probability of the adjacent upward site in the bulk being wet.
Thus   we have the recurrence
\begin{align}
	\bar{S}_{m,m-1}=p\bar{S}_{m,m}+q\bar{S}_{m-1,m-1}+mQ_{m,m-1}\,, \quad m >1\,.
\label{Swallp}
\end{align}
This is in fact similar to the case adjacent to the wall in the dry wall problem~\cite{EssamGutt95}, as this is the point where the cluster growth is restricted. 
Again we impose separately a condition for $m=1$,
\begin{align}
	\bar{S}_{1,0}=p\bar{S}_{1,1}+Q_{1,0}\,.
\label{Swallp1}
\end{align}

\subsubsection{Low density constraints:}
Here we restrict our study of the mean size to the low-density region. In the low-density region there are no infinite clusters, 
so for all of the above equations we will use the fact that
\begin{align}
	Q_{m,y}=1 \ \mbox{ for \ } p<\frac{1}{2}\,,
\label{Qmy1}
\end{align}
as shown for the general damp wall in \cite{Lonsdalebias}. We require that in the limit $y\to \infty$\,, where the cluster is no longer 
affected by the wall, the mean size must behave like the bulk result \cite{Essam89}.
So, for the low-density region of $p<\frac{1}{2}$\,, we have
\begin{align}
	\lim_{y\to \infty} S_{m,y}&=\frac{m}{1-2p} \left(\frac{(1-p)^2}{1-2p}+\frac{m-1}{2} \right)\,.
 \label{Sinf2} 
\end{align}

\subsection{Series expansion for the low-density region}

Using \Ref{Sbulk}--\Ref{Qmy1} we can derive a series expansion for the mean size in the low-density region for a given $m$\/ and $y$. 
For the case of a seed of width one adjacent to the wall, that is $m=1$ and $y=1$, the mean size is equal to 
\begin{align}
	S_{1,1}(p,p_w)
	 = & \  (1 + p_w) 
	 +(1 + 2 p_w + 2 p_w^2)p
	   +(2 + 2 p_w + 3 p_w^2 + 5 p_w^3)p^2
	   \nn & 
	         + (4 + 5 p_w + 3 p_w^2 + 2 p_w^3 + 14 p_w^4)p^3
	   \nn & 
	          + (8 + 8 p_w + 11 p_w^2 + 9 p_w^3 - 14 p_w^4 + 42 p_w^5)p^4
	             \nn & 
	        +  (16 + 19 p_w + 11 p_w^2 + 16 p_w^3 + 58 p_w^4 - 108 p_w^5 +  132 p_w^6)p^5
	   \nn & 
	        +  (32 + 30 p_w + 48 p_w^2 + 26 p_w^3 - 71 p_w^4 + 387 p_w^5 -  561 p_w^6 + 429 p_w^7)p^6
	   \nn & 
	        + O(p^{7})\,.
\label{S11series}
\end{align}
We note that the constant coefficient of $p^n$, for $n\ge 1$, is equal to $2^{n-1}$, which allows us to derive 
the dry wall result given in \Ref{Sdry11}, for $p_w=0$, as a simple geometric series.  
We  further  note the presence of  Catalan numbers in the mean size series expansion above, appearing as the 
coefficient of the highest power of $p_w$\/ for a given power of $p$.

\section{Investigating the mean size}

\subsection{Using a dry wall form of solution}

Guided by the work near a dry wall  \cite{EssamGutt95}, we attempt to solve the recurrence relations for the mean size near a damp wall, 
in the low-density region,  using a similar form of solution.  It was noted in \cite{EssamGutt95} that the mean size in the bulk case, $S_m^{\rm bulk}$,  
is a particular solution to the inhomogeneous part of  \Ref{Sbulk}, and the solutions to the homogeneous part were given in~\cite{EssamTan94}. 
Of the solutions to the homogeneous part we choose only those which vanish as $y\to\infty$, to satisfy~\Ref{Sinf2}. So we try a solution of the form
 \begin{align}
	 S_{m,y}=\left\{
			\begin{matrix}
				S_m^{\rm bulk}(p) + f(m;p,p_w)\lambda^{y-m}\,, & y \ge m-1\,,
				\\ \\
				S_m^{\rm bulk}(p) + g(m;p,p_w)\,,  \qquad & y= m\,, \qquad
				\\ \\
				S_m^{\rm bulk}(p) + h(m;p,p_w)\,, \qquad & y= m+1\,,
			\end{matrix}
			\right. 
\label{Sform}
 \end{align} 
 in the low-density region. In line with the result from the dry case we first used trial functions  $f$, $g$\/ and $h$\/   linear in $m$; 
 however we found that substituting this form of trial solution into  the recurrence relations \Ref{Sbulk}--\Ref{Swallp1} produces an inconsistent system.

Next we attempted to use a form of trial solution allowing $f$, $g$\/ and $h$\/ to be polynomials in $m$\/ of degree up to 2, which is the degree of the bulk mean size, 
but we were again unable to satisfy all recurrences.  This was the case even when the form of solution in \Ref{Sform} was generalised further to a more 
general $y$-dependence, and also if the assumption of the bulk term was removed. As a result we conclude that the mean size in the 
damp wall case must have a different form of solution to the dry wall mean size, and that the solution is of a more complicated form. 
This is perhaps not surprising given the presence of Catalan numbers in the series expansion in  \Ref{S11series},  
which leads us to believe that the generating function for the mean size is not rational.

\subsection{Functional equations}
\label{functional}

A functional equation approach, trying to apply the kernel method, was also unsuccessful. 
This required a two variable generating function to be formed, in variables conjugate to the width of 
the cluster in the left-most column (variable $a$) and the
height of the cluster above the wall (variable $b$).
For the dry wall case, the known solution is equivalent to a rational form for the generating function
with one triple pole in $a$ and two poles in $b$.
However, the kernel method approach is somehow degenerate for the dry wall case
so that the equation has to be solved in a non-standard way, to find the
rational function.

An expansion of the generating function in the damp wall case displays Catalan
numbers: a likely sign that at best the generating function is algebraic and
more probably transcendental.
An approach to this kind of  problem brings us to the cutting edge of
kernel problems and the kernel of the functional equation displays a group
of 8 symmetry. Thus there are eight different transformations of the catalytic variables that leave the kernel invariant and this can, in principle, be used to find a solution. However,
because of the complicated nature of the coefficients in the functional
equation it will only lead to an algorithm and not a closed form.

\subsection{Seeking recurrences for coefficients:  \boldmath $p_w=2p$}\label{pw2p}

Since the method used near a dry wall for the mean size was not successful for the damp wall, and the kernel method did not work, 
we must try other methods. Rather than working directly with the recurrence relations, we consider working with a series expansion 
for the mean size produced from them instead.

The series expansion in \Ref{S11series} is difficult to analyse; no relationship could be found for the coefficients of $p$ 
in terms of $p_w$, or vice versa. It is preferable to consider a simpler case which would  result in a series in a single variable, 
ideally with integer coefficients. This can be achieved by setting $p_w$ equal to an integer; however, the two integer values 
in the domain of $p_w$\/ correspond to special cases --- with $p_w=0$ corresponding to a dry wall and $p_w=1$ to a wet wall. 
Instead we set $p_w$\/ equal to an integer multiple of $p$, as this will also lead to a series in $p$ alone with integer coefficients. 
The case $p_w=p$\/ is also a special case, the neutral wall which is equivalent to a variant of the dry wall scenario 
as noted in \cite{Lonsdale11}. We thus choose to work with $p_w=2p$, and note that for the low-density region 
$p<\frac{1}{2}$ this will remain in the domain of $p_w$\,.

The first few terms in the low-density expansion of $S_{1,1}$ with $p_w=2p$, obtained from recursively generating  \Ref{Sbulk}--\Ref{Qmy1}, are
\begin{align}
S_{1,1}(p,2p) = & \  1 + 3p + 6p^2 + 16p^3 + 30p^4 + 84p^5 + 130p^6 + 464p^7 +  380p^8  
\nn &
+ 3048p^{9}-1666p^{10}+27232p^{11}  -60116p^{12} + 332216p^{13} + O(p^{14})
\label{seriesS2p}
\end{align}
We used the Guess.m package \cite{guess} for Mathematica to analyse the first fifty terms of the series for $S_{1,1}(p,2p)$, 
and found that the coefficients satisfy the recurrence relation
\begin{align}
	 (  n + 2 )^2a_n - (  n + 2 )( 3n + 7) a_{n-1} - 2( 7n^2 - 20n +  4 )a_{n-2}
		  & \nn
	  +  16( 5n^2 -15n  +13)a_{n-3} - 16( 9n^2 -40n  +46 )a_{n-4} 
		& \nn
	+ 16( n-3 )( 7n - 22) a_{n-5} - 32( n - 4)^2 a_{n-6}
	   &=0\,,
   \label{ansizerec}
\end{align}
where $a_n$\/ is the $n^{th}$\/ coefficient of the mean size. That is, we make the definition
\begin{align}
	S:= S_{1,1}(p,2p)=\sum_{n=0}^\infty a_np^n\,.
\label{San}
\end{align}
From the recurrence in \Ref{ansizerec}, we can find a second order inhomogeneous differential equation satisfied by $S$, 
\begin{align}
 2(1-2p)(2-11p-14p^2+76p^3-88p^4+32p^5)S &
\nn  + \ p(1-2p)^2(5-2p-58p^2+96p^3-40p^4)S' &
 \nn
 + \ p^2(1-p)(1-2p)^3(1+4p -4p^2)S'' 
= & \ 4-3p-44p^2+86p^3-72p^4+24p^5\,,
\end{align}
which has a confluent singularity at $p=\frac{1}2$\, with exponents 2 and $-2$.
So we can conclude that for $p_w=2p$\/ the critical exponent $\gamma=2$, which equals the exponent  found in the wet wall case.
This follows what we might expect, as for $p_w=2p$\/ letting $p\to \frac{1}2$\, we approach the wet case of $p_w=1$. 
So this corresponds to another ``special case'' value of $p_w$\,; although it is  not exactly the same as the wet wall scenario, it will be equivalent at the critical point.  

We attempted to apply the same method for other $p_w=kp$, where $k$\/ is an integer, but  were unable to 
find any other simple recurrences for the mean size  and  therefore looked to other methods of analysing these cases.

\section{Modular arithmetic method}\label{modularsection}

We used a series analysis method, outlined in  \cite{BGHJ_2008} and \cite{Jensen2010},  to determine the minimal order linear homogeneous
ODE satisfied by the mean size, and from  this derive the critical behaviour. For convenience we will simply refer to this method   as 
``the modular arithmetic method'',  since the computation of the ODE  is performed modulo specific primes.

\subsection{The general case  \boldmath $p_w=rp$}

We consider the case where the wall occupancy probability $p_w=rp$, where  $r$\/ is a rational number.  
The series expansion for the mean size in this case is 
\begin{align}
S_{1,1}(p,rp) \ 
 = & \  1 
 +(1 + r)p
   +(2 +   2 r  )p^2
     + (4 +2r+ 2 r^2)p^3
          + (8 +5r+ 3r^2  )p^4
             \nn & 
        +  (16 + 8r+ 3 r^2+5 r^3 )p^5
        +  (32  +19 r+ 11 r^2 + 2 r^3  )p^6
   \nn & 
        + (64 + 30 r+ 11 r^2 + 9 r^3 + 14 r^4)p^7
        + O(p^{8})\,.
\label{S11prp_series}
\end{align}
Applying the modular arithmetic method  \cite{BGHJ_2008,Jensen2010}   to this series,  we were able to reconstruct the 
minimal ODEs satisfied by the mean size for any given $r$. 

We found that, for general rational values of $r$, the series for the mean size is a solution to a homogeneous ODE of order 4 and degree 33.
The exceptions are $r=2$, which reduces to a third order ODE, and also  $r=1$ and 0, which both reduce to first order ODEs. 

\subsection{Singularities}\label{singularities}
We locate the singularities of the problem by analysing the {\it head polynomial}, that is the polynomial coefficient of 
the highest order term in the minimal ODE.  For a given value of $r$, where $p_w=rp$, we generate and factorise the head polynomial, which was in general of degree 33. 
We remove  factors corresponding to apparent singularities, in the form of high degree polynomials, and work with the remaining factorised polynomial, generally of degree 13.  For the examples $r=3,4,5$ and 6, this tells us that the singularities are given by the roots of the following polynomials: 
\begin{align}
	Q_4(p,3p) & = (1-2p)^4(1-p)(1+4p-4p^2)(1-3p)(2-3p)(1-18p^2+36p^3-18p^4)\,;
\label{Q4_3p}
\\
	Q_4(p,4p) & = (1-2p)^4(1-p)(1+4p-4p^2)(1-4p)(3-4p) (3 -64p^2+128p^3-64p^4)\,;
\label{Q4_4p}\\
	Q_4(p,5p) & = (1-2p)^4(1-p)(1+4p-4p^2)(1-5p)(4-5p) (1-25p^2+50p^3-25p^4)\,;
 \label{Q4_5p}
\\
	Q_4(p,6p) & = (1-2p)^4(1-p)(1+4p-4p^2)(1-6p)(5-6p) (5-144p^2+288p^3-144p^4).
 \label{Q4_6p}
\end{align}
We then seek to write a general expression for the head polynomial in terms of $r$\/ and $p$\/ --- 
which is why we have chosen integer values of $r$, so that we may easily generalise our result using these expressions. 
Naturally we have also avoided the special cases of $r=0,1$ and 2, so that we may find the general damp behaviour 
without considering the isolated exceptions which behave like the dry or wet wall.

Looking at equations  \Ref{Q4_3p}--\Ref{Q4_6p},  the linear factors can be easily determined in terms of $r$, 
and in fact some factors are constant for all $r$.  For the quartic factor  we utilise that any ODE can be multiplied by an arbitrary constant. 
Based on the pattern observed we make the guess that the constant term of the quartic is equal to $r-1$.
With this ansatz it easily follows that the other coefficients are given by $-4r^2$, $8r^2$ and $-4r^2$, 
respectively, and so we have determined the general form of the quartic.  
Thus we find that in the general case the singularities are given by the roots of the following polynomial
\begin{align}
	Q_4(p,rp)&= (1-2p)^4(1-p)(1+4p-4p^2)  (1-rp) \left(r-1-r p\right)P_4(p,rp), \nonumber \\  
	\mbox{where} \quad   P_4(p,rp)&=r-1-4r^2p^2+8r^2p^3-4r^2p^4\,. \label{P4_rp}
\end{align}
Although we have calculated this using integer values of $r$, it can be verified that this holds for any rational $r$. 
Since the method used to find the ODE assumes that the series coefficients (and hence the coefficients in the ODE) are integers modulo a prime number, this cannot be directly extended to irrational values of $r$, but it is reasonable to expect that the behaviour when $p_w=rp$ for $r$ real would be the same as that for $r$ rational.

\subsubsection{Roots of $Q_4(p,rp)$:}

The roots of the quartic, $P_4(p,rp)$, are
\begin{eqnarray}
	p_{4,1}=\frac{r+\sqrt{r^2+2r\sqrt{r-1}}}{2r}, && p_{4,2}=\frac{r-\sqrt{r^2+2r\sqrt{r-1}}}{2r}, \nonumber \\
	p_{4,3}=\frac{r+\sqrt{r^2-2r\sqrt{r-1}}}{2r}, && p_{4,4}=\frac{r-\sqrt{r^2-2r\sqrt{r-1}}}{2r}.
\end{eqnarray}
which combine with the other roots of $Q_4(p,rp)$, which are 
\begin{eqnarray}
	p_{1}=\frac{1}{2}, \ \ p_{2}=1, \ \ p_3^{\pm}=\frac12 \big(1\pm \sqrt{2}\big), \ \ 
	p_{4}=\frac{1}{r}, \ \ p_{5}=1-\frac{1}{r}.
\end{eqnarray}
and so we have  the singularities for the general damp case in terms of $r$. The associated critical exponents, found using Maple to solve the indicial equation of the ODE at the singularity, are listed in Table~\ref{tab:exp_gen}.

\begin{table}[htdp]
\caption{Critical exponents at the roots of  $Q_4(p,rp)$ \label{tab:exp_gen}}
	\begin{center}
	\begin{tabular}{lr}
		\hline \hline
		Singularity \quad \quad & \multicolumn{1}{l}{Exponents} \\
		\hline
		 $0$ & $-2,\,\, -2,\,\,\, -1,\quad \, 0$ \\
		 $ 1/2$ & $-1,\quad \, 1,\quad \,1,\quad \,3$ \\
		 $ (1\pm \sqrt{2})/2$ & $0,\quad \, 1,\quad \,2,\quad \, 4$ \\
		$  1/r$ & $-2,\quad \, 0,\quad \, 1,\quad \, 2$ \\
		$  1-1/r$ & $-3,\quad \,  0,\quad \, 1,\quad \,  2$ \\
		 $1$ & $0,\quad \,   0, \quad \, 1,\quad \, 2$\\
		 $P_4(p,rp)$ &$1/2,\quad \,  0,\quad \, 1,\quad \,  2$ \\
		 $\infty$ & $1,\quad \,  2,\quad \, 2,\quad \,  4$\\
		\hline \hline
	\end{tabular}
	\end{center}
\label{exponents_Q4}
\end{table}%

We consider the singularities for different values of $r$. 
Since $p_w=rp$\/ is a probability, and we are looking at the low-density region $p<\frac{1}{2}$, we will consider $r\in[0,2]$.
We recall that $r=0$ and $r=1$ correspond to dry and `dry-like' cases, while $r=2$ corresponds to a `wet-like' case. Hence it is natural to consider the behaviour between these exceptions.

In the region $0<r<1$ the closest singularity on the positive real axis is $p_c=1/2$, while
for $4-2\sqrt{3} <r<1$ the pair of complex conjugate roots $p_{4,2}$ and $p_{4,4}$ of  $P_4(p,rp)$ are closer to the origin.
In the region $1<r<2$ the closest singularity on the positive real axis is $p_c=1-1/r$, though
the negative root $p_{4,2}$ from $P_4(p,rp)$ is closer to the origin after $r>1.186659\dots$. 
It is of interest to note that when $r=2$, three of the singularities in Table \ref{exponents_Q4} coalesce at $p_c=\frac{1}{2}$, highlighting this special case.

\subsubsection{Analysis of singularities:}

\begin{figure}
\begin{center}
\includegraphics[width=14cm]{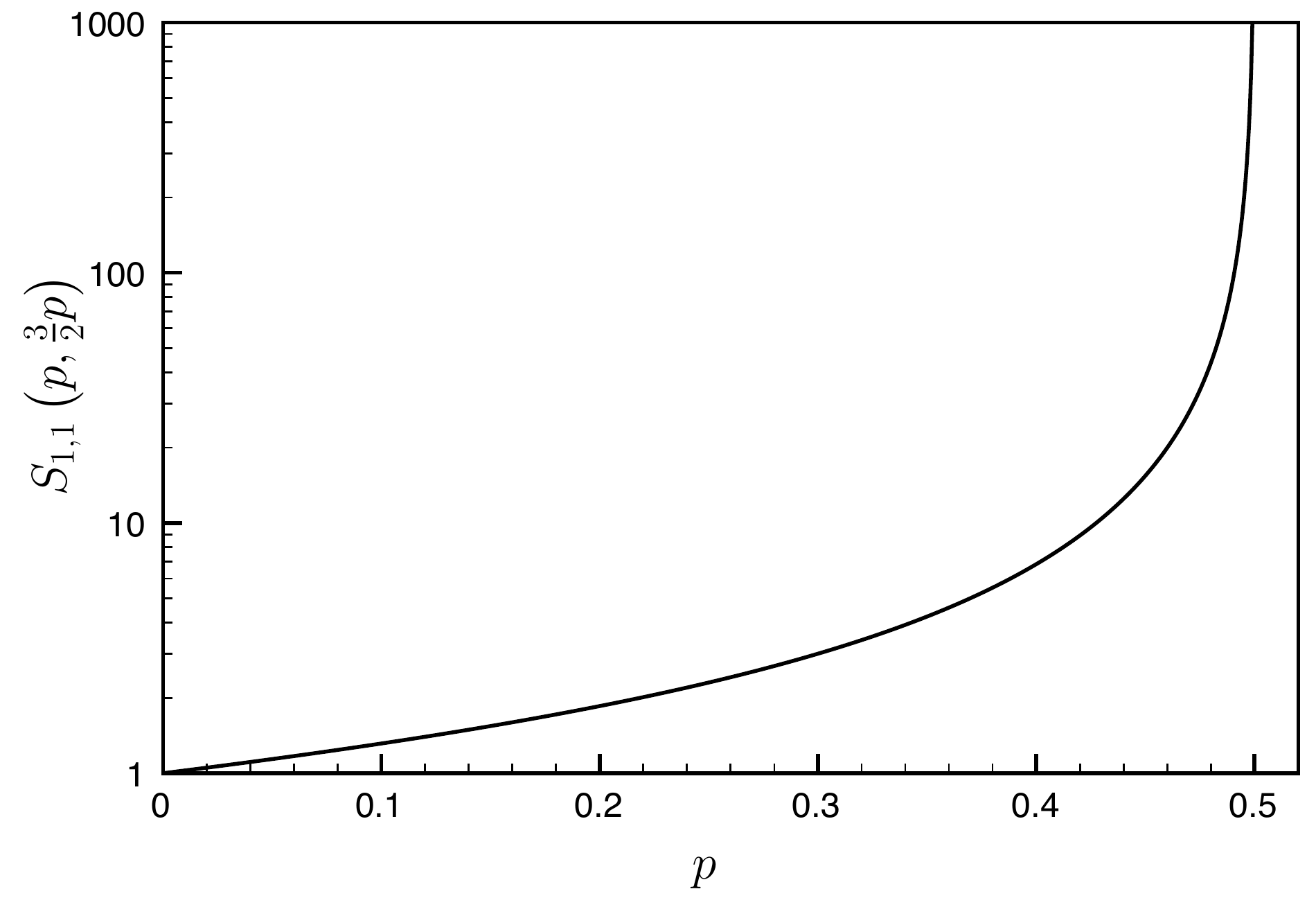}
\end{center}
\caption{\label{fig:size32} The mean size in the low-density region for $r=\frac32$.}
\end{figure}

The roots of $Q_4(p,rp)$ give an indication of the singularities of the mean size, when we consider only the positive real axis. 
For $0<r<1$ we are thus not surprised to see a singularity at $p_c=1/2$, as this is the critical point for directed compact percolation. 
However, the singularity  $p_c=1-1/r$\/ found in the region $1<r<2$ looks suspect on physical grounds.
It is hard to imagine how a damp wall could lead to a physical critical point that is lower than the one
for the wet case. And we shall indeed see that while $1-1/r$\/ is a singularity of the ODE, and appears
explicitly in exact solutions to the ODE, the actual physical low-density series is {\em not}\/ singular at
$1-1/r$\/ but the physical singularity occurs at $p_c=1/2$. A simple numerical demonstration will suffice.
We take the low-density series, which is correct to order 200 in $p$ for any rational value of $r$, 
and look at  a Pad\'e approximant (ratio of polynomials) to $S_{1,1}(p,rp)$ with $r$ fixed. Plotting the Pad\'e approximants
as function of $p$ clearly shows that the series is not singular at $1-1/r$, and only diverges at $p=1/2$.
In Fig.~\ref{fig:size32} we plot a Pad\'e approximant with degree 50 polynomials to $S_{1,1}(p,\frac32 p)$.
Clearly there is absolutely no sign of a singularity at $p=1/3$ (note that the critical exponent at $1-1/r$ is $-3$
so we should see  divergence if the series was singular). 
We note that in the mean length calculation~\cite{EssamDomb} a similar apparent divergence was found at $q=1/(1+p_w)$. 
With the choice of  $p_w=rp$, the specific value of $p=1-1/r$\/ satisfies $q=1/(1+p_w)$, and so this is the same mathematical quirk.

Naturally we are  interested in the critical exponent for the percolation problem, so we focus on $p_c=\frac{1}{2}$,
which is the physical singularity in the damp region $0<r<2$.
The dominant behaviour is divergence with an exponent of 1, which corresponds to the exponent 
$\gamma^{\rm damp}=1$, in line with the dry wall mean size. 
When $r=2$ we tend to a wet wall scenario and we have $\gamma=2$.

\subsection{Solutions to ODE}

We further examined the ODEs using the very powerful Maple package {\tt DETools}. Trying to solve a given ODE with {\tt dsolve} yielded for each value of $r$ a  simple algebraic solution, which we show for the examples $r=3,4,5$,
\begin{align}
	{\mathcal S}(p,3p)&=\frac {\sqrt {1-18\, p^2+36\, p^3-18\, p^4} \left( 1-2\,p \right)  \left( 2+9\,p-9\, p^2 \right) }{ p^2\left( 2-3\,p \right) ^3 \left( 1-3\,p \right) ^2} \,;\\ 
	 \nonumber \\
	{\mathcal S}(p,4p)&= \frac {\sqrt {3-64\, p^2+128\, p^3-64\, p^4} \left( 1-2\,p \right)  \left( 3+16\,p-16\, p^2 \right) }{ p^2\left( 3-4\,p \right) ^3 \left( 1-4\,p \right) ^2} \,; \\ 
	  \nonumber \\
	{\mathcal S}(p,5p) &=\frac {\sqrt {1-25\, p^2+50\, p^3-25\, p^4} \left( 1-2\,p \right)  \left( 4+25\,p-25\, p^2 \right) }{ p^2 \left( 4-5\,p \right) ^3\left( 1-5\,p \right) ^2} \,.
\end{align}
From these, all factors except the quadratic on the numerator are able to be generalised from our previous work on the head polynomial, and the quadratic factor is easily expressed in terms of $r$. 
Hence we can write ${\mathcal S}(p,rp)$ generally as 
\begin{equation}
	{\mathcal S}(p,rp)=
	\frac {\sqrt {{\rm sign}(r\!-1\!)(r\!-\!1-4r^2p^2+8r^2p^3-4r^2p^4)} \left( 1-2p \right)  \left( r\!-\!1+r^2p-r^2p^2 \right) }
	{ p^2\left( r\!-\!1-rp \right) ^3 \left( 1-rp \right) ^2},
\end{equation}
where the ${\rm sign}(r\!-1\!)$ ensures that the square-root has a Taylor expansion with real coefficients. We see that this solution vanishes at $p=p_c=\frac{1}{2}$.

\subsubsection{Rational solution:}
The Maple package {\tt DETools} also has a number of procedures to look for simple solutions and one of these, {\tt ratsols}, 
yielded a rational solution ${\mathcal R}(p,rp)$ for any value of $r$. We were particularly interested in the region   $1<r<2$, 
as this  spans between the two special cases of the `dry-like' neutral wall, $p_w=p$, and the `wet-like' case of $p_w=2p$.
It turns out that in this region  the rational solution is the dominant behaviour.

However, for the sake of generalising the solution, we again focus on integer values of $r$\/ other than the special cases. 
 For $r=3$ we have a rational solution, equal to 
\begin{equation}
	 {\mathcal R}(p,3p)=
	 \frac {-10+43\,p-87\,{p}^{2}+596\,{p}^{3}-2688\,{p}^{4}+5292\,{p}^{5}-4752\,{p}^{6}+1620\,{p}^{7}}
	 {p^2 \left( 1-2\,p \right)  \left( 1-3\,p \right) ^{2} \left( 2-3\,p \right) ^{3}}\,.
\end{equation}
From this, the denominator can be seen to follow the pattern of factors already found in Section \ref{singularities}.
However, the numerator $N(p,rp)$ is a little more tricky. 
Looking at it, for $r=3,4,5$, we have
\begin{align}
	N(p,3p)&=-10+43p-87p^2+596p^3-2688p^4+5292p^5-4752p^6+1620p^7, \\  
	N(p,4p)&=-9+24p+3p^2+543p^3-3144p^4+6304p^5-5504p^6+1792p^7,   \\  
	N(p,5p)&=-44+61p+311p^2+3228p^3-20720p^4+41450p^5-35500p^6+11250p^7. 
 \end{align}
There is no clear pattern at this stage, but we have not yet utilised the arbitrary constant that can simplify the search for a general expression. 
At first we tried to express the coefficients of $N(p,rp)$ as a polynomial in $r$; however, with an arbitrary constant multiplying each 
$N(p,rp)$, this is an ill-defined problem. Hence we need to somehow determine the arbitrary constant.
 
We note that  since  ${\mathcal R}(p,rp)$ is a solution of the ODE, it is possible that it appears as part of a direct sum 
decomposition and can be `removed' from $S_{1,1}(p,rp)$. That is, we form the function
 \begin{align}
	 G_r(p)=p^2S_{1,1}(p,rp)-c_rp^2{\mathcal R}(p,rp)
 \end{align}
noting that the extra factor $p^2$\/ is introduced to cancel the $1/p^2$ appearing in ${\mathcal R}(p,rp)$. 
Generically, $G_r(p)$ will be a solution only to the original fourth order ODE\null. 
However, if the rational solution is  removable then there is a unique value of $c_r$\/ for which $G_r(p)$ 
becomes a solution of a third order ODE\null.  The value of $c_r$\/ must be rational since both functions 
have rational Taylor coefficients.  So we form the series for $G_r(p)$ modulo a prime and do a brute force 
search through all values of $c_r$\/ up to the value of the prime. We search for an ODE of order 4 and degree 33 
as  originally done, which normally requires 170 series terms.  For a particular value of $c_r$\/  far fewer terms 
are needed, signifying that for this value the ODE simplifies to a third order ODE 
(sometimes two primes were required to determine $c_r$\/ uniquely). We did this for integer values of $r=3,\dots,10$, 
and thus formed the polynomials $c_rN(p,rp)$. We found that the coefficients of these polynomials could be 
expressed as polynomials in $r$ of degree at most 5, as follows: 
\begin{eqnarray}
	a_0(r)&=& 4-11\,r+10\,{r}^{2}-3\,{r}^{3},    \\ 
	a_1(r)&=&-16+52\,r-60\,{r}^{2}+27\,{r}^{3}-3\,{r}^{4}  \\ 
	a_2(r)&=& 24-96\,r+148\,{r}^{2}-91\,{r}^{3}+15\,{r}^{4},  \\ 
	a_3(r)&=&-8+64\,r-176\,{r}^{2}+156\,{r}^{3}-44\,{r}^{4}+8\,{r}^{5} \\ 
	a_4(r)&=& -16\,r+128\,{r}^{2}-168\,{r}^{3}+96\,{r}^{4}-40\,{r}^{5}, \\ 
	 a_5(r)&=&-48\,{r}^{2}+96\,{r}^{3}-112\,{r}^{4}+72\,{r}^{5} \\
	 a_6(r)&=& -16\,{r}^{3}+56\,{r}^{4}-56\,{r}^{5}, \\ 
	  a_7(r)&=&-8\,{r}^{4}+16\,{r}^{5}.  
\end{eqnarray}

\subsection{Factorising the differential operator}
 
Following from these results, we naturally conjecture that the fourth order differential operator $L_4(p,rp)$ 
for $S_{1,1}(p,rp)$ can be written as a product of a second order operator  $L_2(p,rp)$ and two first order operators  
$L_{\mathcal{S}}(p,rp)$ (for the square-root type solution) and  $L_{\mathcal{R}}(p,rp)$  (for the rational solution) 
with the latter appearing as part of a direct sum decomposition, 
 \begin{equation}
	L_4(p,rp)= L_2(p,rp)\cdot  L_{\mathcal{S}}(p,rp) \oplus L_{\mathcal{R}}(p,rp)\,.
\end{equation}
As we have already investigated $\mathcal{S}$ and $\mathcal{R}$, we  now turn to the determination of  $L_2(p,rp)$.
 This involves  finding the polynomials $Q_j(p,rp)$ such that
 \begin{align}
	  L_2(p,rp) = Q_2(p,rp)\frac{\mathrm{d} ^2}{ \mathrm{d} p^2}+Q_1(p,rp)\frac{\mathrm{d} }{\mathrm{d}  p}+Q_0(p,rp) \label{L2Qs}\,.
  \end{align} 
To do this we make use of some very powerful procedures  from {\tt DETools}. First we calculated the series
expansion for $S_{1,1}(p,p_w)$ to order 200 in $p$ and order 100 in $p_w$ yielding series correct to
order 200 in $p$ for any value of $p_w=rp$. We then fixed $r$ at an integer value $\geq 3$ and 
used our ODE finder to calculate $L_4(p,rp)$ modulo several primes. From these modular results
we then reconstructed the exact ODE, which required the use of 10 distinct primes. 
 Next we used the procedure {\tt DFactorLCLM}  from {\tt DETools}  to factor $L_4(p,rp)$ (for fixed $r$) into a direct sum 
 of a third order operator and $L_{\mathcal{R}}(p,rp)$, and next we use the procedure {\tt DFactor}  to factor 
 out $ L_{\mathcal{S}}(p,rp)$ leaving us with $L_2(p,rp)$. For $r=3$ we find that
\begin{align}
	Q_2(p,3p)&= p^2(1\!-\!2p )(1\!-\!p)  (1\!+\!4p\!-\!4p^2  )   ( 2\!-\!3p )
	 (  1\!-\!18p^2\!+\!36p^3\!-\!18p^4 )\times   \nonumber \\
	& \qquad \quad   (  2\!+\!9p\!-\!9{p}^2  ) ^2 ( 1\!-\!2p\!+\!11p^2\!-\!90p^3\!+\!225p^4\!-\!216p^5\!+\!72p^6)\,.
 \end{align}
All factors except one are known from our previous working, and we just need to find the remaining sixth degree polynomial. For $r=4$ this polynomial is
\begin{align}
	3-6p+38p^2-320p^3+800p^4-768p^5+256p^6
\end{align}
Again keeping in mind the presence of an arbitrary constant, and the previous results, it seems likely that the constant term is  just $r-1$, which helps us fix the `normalisation' --- and indeed we find that the general expression for the sought after polynomial is
\begin{align}
	r-1-2( r-1) p-2( 1-r-r^2 ) p^2-20r^2p^3+50r^2p^4-48r^2p^5+16r^2p^6\,,
\end{align}
so that 
\begin{align}
	Q_2(p,rp)&=  p^2(1-2p)(1-p) (1+4p-4p^2) (r-1-rp) \times  \nonumber \\ 
	& \qquad (r-1+r^2p-r^2p^2)^2 ( r-1-4r^2p^2 +8r^2p^3- 4r^2p^4)\times \\
	 & \qquad \Big(r\!-\!1-2\left( r\!-\!1 \right) p-2 \left( 1\!-\!r\!-\!r^2 \right) p^2\!-\!20r^2p^3 \!+\!50r^2p^4\!-\!48r^2p^5\!+\!16r^2p^6\Big).  \nonumber
 \end{align}
Similarly, although more cumbersome to compute, we found expressions for $Q_1(p,rp)$ and  $Q_0(p,rp)$ --- 
the full polynomials are listed in \ref{app:poly}.
Thus we found the second order operator  $L_2(p,rp)$ as defined in  \Ref{L2Qs}. 
To accomplish this feat the calculations outlined above were repeated for all integers $r$ between 3 and 14
and the resulting expressions for $L_2(p,rp)$ were used to calculate the general expressions 
for  $Q_1(p,rp)$ and  $Q_0(p,rp)$. As a technical aside this again required us to `fix' an arbitrary
constant say by fixing the very simple expressions for the highest degree coefficients.

\section{Results and conclusions}

We were not able to investigate the mean size of directed compact percolation near a damp wall using methods successful in finding exact closed form
solutions for the dry wall case.  However, we did have some success with analysing the series expansion. 
Using the modular arithmetic method we found an ODE for any rational $r$\/ where $p_w=rp$, and these ODEs 
completely determine the singular behaviour of the system. So even though we were not able to find a closed form solution, 
we are still able to analyse the result for mean size near a damp wall.

\subsection{Differential equation}

For wall occupancy probability $p_w=rp$, where $r$\/ is a rational number, 
the series for the mean size is a solution to an homogeneous ODE of order 4 and degree 33. 
 The exceptions are for $r=2$, which reduces to a third order ODE, and $r=1$ and 0, which both reduce to first order ODEs.

The $r=0$ and $r=1$ cases correspond to the dry and neutral wall mappings, $p_w=0$ and $p_w=p$\/ respectively, 
and thus it is not surprising that we have a greatly simplified situation for these values of $p_w$\,.
The $p_w=2p$\/ case is a special `wet-like' case that we considered with our initial series analysis techniques 
in Section~\ref{pw2p}, and we found it to be satisfied by an inhomogeneous ODE of order 2.
There is no contradiction between this result and our later finding that the minimal order ODE for the 
$p_w=2p$\/ case is three, because this refers to  homogeneous ODEs of Fuschian type.

In the region $1<r<2$ the ODE has a singularity at $1-1/r < 1/2$, which at first sight  appears to be the
physical singularity. However, the simple numerical test of looking at Pad\'e approximants to the
actual series shows that the series itself is not singular until $p_c=1/2$. So in the whole of the damp
region the physical singularity occurs at $p_c=1/2$. The singularity at  $1-1/r$ is explicitly present
in the two particular solutions we found to the ODE so somehow the solutions combine in such
a way as to cancel this singularity in the physically relevant linear combination yielding
the low-density series.

\subsection{Critical exponent}

Although we do not have an expression for the mean size, the ODEs give us the critical behaviour. 
So we find that for $p_w=rp,$ where $0\le r<2$, the critical exponent 
\begin{align}
	\gamma^{\rm damp}=1\,,
\end{align}
which is the same as the dry wall exponent. 
This includes the special cases $p_w=0$ and $p_w=p$, which both correspond exactly to a dry wall scenario.
For $p_w=2p$\/ we find that the exponent $\gamma=2$, the same as the wet wall exponent. 
This case is not exactly a wet wall situation, but as $p\to\frac{1}{2}$\, the proportion of wet sites on the wall tends to 1, so it is `wet-like'.
We note that although we have only considered the general damp case for $p_w=rp$, the finding of a common critical exponent 
for all $0<r< 2$ implies that this should be the same for all $p_w$\/ except for `wet-like' cases. 
These exponents are summarised in Table \ref{meansizeexponents}.

\begin{table}[htdp]
\begin{center}\begin{tabular}{|c|c|c|c|c|c|}
	\hline
	$p_w$ & 0 & $p$ & $rp$ & $2p$ & 1  \\
	\hline
	$\gamma$ &  1 & 1 & 1 & 2 & 2 \\
	\hline
	\end{tabular} 
\caption[Critical exponents for mean size, for different values of $p_w$]{Critical exponents for mean size, for different values of $p_w$\,. 
Note that for $p_w=rp$\/ we have listed the result for $0<r< 2$.} \label{meansizeexponents}
\end{center}
\end{table}

Thus, as  expected, the mean size follows the pattern of the percolation probability, mean length and mean number of contacts, 
in that its critical behaviour is the same as the corresponding dry wall result. 
We have  the dry wall exponent $\gamma=1$ for $p_w\in[0,1)$, and the wet exponent  
$\gamma=2$ only  at the wet singularity $p_w=1$, and at the wet-like case of $p_w=2p$.

\section*{Acknowledgements}
The authors would like to thank Andrew Rechnitzer for his assistance with the functional equation approach outlined in Section \ref{functional}.

Financial support from the Australian Research Council via its support for the Centre of Excellence for Mathematics and Statistics of Complex Systems 
is gratefully acknowledged by the authors. IJ and ALO were supported by funding under the Australian Research Council's Discovery Projects scheme 
by the grant DP120101593.


\appendix
\section{The Polynomials $Q_1(p,rp)$ and  $Q_0(p,rp)$ }
\label{app:poly}

\subsection{$Q_1(p,rp)$ }

\begin{align}
Q_1(p,rp)=p(r-1+r^2p-r^2p^2)\sum_{n=0}^{17} b_n(r)p^n
\end{align}
where

\begin{gather}
\begin{aligned}
b_0(r) & = 6(r-1)^4 \\
b_1(r) & = (r-1)^3(r-2)(6-7r) \\
b_2(r) & = -(r-1)^2(28-42\,r-39\,{r}^{2}+62\,{r}^{3}) \\
b_3(r) & = (r-1)^2 (112-280\,r+104\,{r}^{2}+125\,{r}^{3}-34\,{r}^{4}) \\
b_4(r) & = 2 (r -1) (70-316\,r+685\,{r}^{2}-689\,{r}^{3}+221\,{r}^{4}+48\,{r}^{5}) \\
b_5(r) & =  -2(r-1) (40-266\,r+1808\,{r}^{2}-3351\,{r}^{3}+2266\,{r}^{4}-480\,{r}^{5}+36\,{r}^{6}) \\
b_6(r) & = -2(8-128\,r+3836\,{r}^{2}-12328\,{r}^{3}+16107\,{r}^{4}-10619\,{r}^{5}+3568\,{r}^{6}-488\,{r}^{7})\\
b_7(r) & = -2 r (24-5336\,r+19108\,{r}^{2}-25748\,{r}^{3}+18075\,{r}^{4}-7763\,{r}^{5}+2180\,{r}^{6}) \\
b_8(r) & = -4 r^2 (2372-9376\,r+8729\,{r}^{2}+261\,{r}^{3}-2058\,{r}^{4}-872\,{r}^{5}) \\
b_9(r) & = 4 r^2 (1168-5500\,r-7460\,{r}^{2}+33837\,{r}^{3}-31571\,{r}^{4}+10118\,{r}^{5}) \\
b_{10}(r)& =-8 r^2 (120-856\,r-11826\,{r}^{2}+38080\,{r}^{3}-42551\,{r}^{4}+23886\,{r}^{5}) \\
b_{11}(r)& =-8 r^3 (104+12720\,r-45638\,{r}^{2}+64846\,{r}^{3}-55321\,{r}^{4}) \\
b_{12}(r)& =16 r^4 (3728-16832\,r+31869\,{r}^{2}-40228\,{r}^{3}) \\
b_{13}(r)& =-16 r^4 (1184-7664\,r+20668\,{r}^{2}-39027\,{r}^{3}) \\
b_{14}(r)& =64 r^4 (40-496\,r+2147\,{r}^{2}-6343\,{r}^{3}) \\
b_{15}(r)& =64 r^5 (56-520\,r+2665\,{r}^{2}) \\
b_{16}(r)& =512 r^6 (7 - 82 r) \\
b_{17}(r)& =4608 r^7
\end{aligned}
\end{gather}


\subsection{$Q_0(p,rp)$ }
\begin{align}
Q_0(p,rp)= \sum_{n=0}^{19} c_n(r)p^n
\end{align}
where
\begin{gather}
\begin{aligned}
c_0(r) & = 6(r-1)^5\\
c_1(r) & = (r -1)^4(16-28r+15r^2)   \\
c_2(r) & = 4(r -1)^3 (4-15\,r+42\,{r}^{2}-38\,{r}^{3}+2\,{r}^{4}) \\
c_3(r) & = (r -1)^2 (20-48\,r+356\,{r}^{2}-843\,{r}^{3}+711\,{r}^{4}-211\,{r}^{5})  \\
c_4(r) & = -2(r -1)^2 (16-2\,r-220\,{r}^{2}+198\,{r}^{3}+400\,{r}^{4}-537\,{r}^{5}+44\,{r}^{6}) \\
c_5(r) & =  - (r -1) (16+44\,r-4488\,{r}^{2}+14058\,{r}^{3}-16044\,{r}^{4}+6591\,{r}^{5}+163\,{r}^{6}-462\,{r}^{7}) \\
c_6(r) & = -8(r-1)r (8-1482\,r+5311\,{r}^{2}-8213\,{r}^{3}+6705\,{r}^{4}-2631\,{r}^{5}+257\,{r}^{6}-16\,{r}^{7}) \\
c_7(r) & =4 r (4-4152\,r+21289\,{r}^{2}-51652\,{r}^{3}+74765\,{r}^{4}-64156\,{r}^{5}+30124\,{r}^{6}-6658\,{r}^{7}+486\,{r}^{8}) \\
c_8(r) & = 4 r^2 (3312-20040\,r+61048\,{r}^{2}-112851\,{r}^{3}+121460\,{r}^{4}-75705\,{r}^{5}+25080\,{r}^{6}-2980\,{r}^{7}) \\
c_9(r) & = -2 r^2 (2848-22768\,r+90080\,{r}^{2}-206662\,{r}^{3}+258498\,{r}^{4}-196969\,{r}^{5}+86149\,{r}^{6}-17868\,{r}^{7}) \\
c_{10}(r)& = 16 r^2 (64-904\,r+4652\,{r}^{2}-11542\,{r}^{3}+11019\,{r}^{4}-4029\,{r}^{5}+1336\,{r}^{6}-2180\,{r}^{7}) \\
c_{11}(r)& = 4 r^3 (496-2240\,r-9776\,{r}^{2}+82392\,{r}^{3}-186525\,{r}^{4}+139737\,{r}^{5}-31298\,{r}^{6}) \\
c_{12}(r)& = -16 r^4 (304-6836\,r+36692\,{r}^{2}-94962\,{r}^{3}+84274\,{r}^{4}-36391\,{r}^{5}) \\
c_{13}(r)& = 8 r^4 (192-8520\,r+58712\,{r}^{2}-204132\,{r}^{3}+217558\,{r}^{4}-151819\,{r}^{5}) \\
c_{14}(r)& =32 r^5 (640-6752\,r+34336\,{r}^{2}-44638\,{r}^{3}+49757\,{r}^{4})  \\
c_{15}(r)& =-16 r^5 (160-3456\,r+29052\,{r}^{2}-47800\,{r}^{3}+87825\,{r}^{4}) \\
c_{16}(r)& =- 64 r^6 (96-1776\,r+4036\,{r}^{2}-13105\,{r}^{3}) \\
c_{17}(r)& = -64 r^7(192-776\,r+5095\,{r}^{2} ) \\
c_{18}(r)& = -1024r^8(4-73r) \\
c_{19}(r)& = -7680r^9
\end{aligned}
\end{gather}

\end{document}